\pdfoutput=1
\documentclass[12pt,a4paper,titlepage]{article}
\usepackage[utf8]{inputenc}
\usepackage[top=50pt,bottom=50pt,left=68pt,right=66pt]{geometry}
\usepackage{amsmath}
\usepackage{booktabs}
\usepackage{amssymb}
\usepackage[title]{appendix}
\usepackage{mathrsfs}
\usepackage{float}
\usepackage{multirow}
\usepackage{graphicx,caption,subcaption}
\usepackage[space]{grffile}

\begin{document}
\renewcommand{\arraystretch}{1.3}

\makeatletter
\def\@hangfrom#1{\setbox\@tempboxa\hbox{{#1}}%
      \hangindent 0pt
      \noindent\box\@tempboxa}
\makeatother


\def\un#1{\relax\ifmmode\@@underline#1\else
        $\@@underline{\hbox{#1}}$\relax\fi}


\let\under=\unt                 
\let\ced=\ce                    
\let\du=\du                     
\let\um=\Hu                     
\let\sll=\lp                    
\let\Sll=\Lp                    
\let\slo=\os                    
\let\Slo=\Os                    
\let\tie=\ta                    
\let\br=\ub                     


\def\a{\alpha}
\def\b{\beta}
\def\c{\chi}
\def\d{\delta}
\def\e{\epsilon}
\def\f{\phi}
\def\g{\gamma}
\def\h{\eta}
\def\i{\iota}
\def\j{\psi}
\def\k{\kappa}
\def\l{\lambda}
\def\m{\mu}
\def\n{\nu}
\def\o{\omega}
\def\p{\pi}
\def\q{\theta}
\def\r{\rho}
\def\s{\sigma}
\def\t{\tau}
\def\u{\upsilon}
\def\x{\xi}
\def\z{\zeta}
\def\D{\Delta}
\def\F{\Phi}
\def\G{\Gamma}
\def\J{\Psi}
\def\L{\Lambda}
\def\O{\Omega}
\def\P{\Pi}
\def\Q{\Theta}
\def\S{\Sigma}
\def\U{\Upsilon}
\def\X{\Xi}


\def\ve{\varepsilon}
\def\vf{\varphi}
\def\vr{\varrho}
\def\vs{\varsigma}
\def\vq{\vartheta}


\def\ca{{\cal A}}
\def\cb{{\cal B}}
\def\cc{{\cal C}}
\def\cd{{\cal D}}
\def\ce{{\cal E}}
\def\cf{{\cal F}}
\def\cg{{\cal G}}
\def\ch{{\cal H}}
\def\ci{{\cal I}}
\def\cj{{\cal J}}
\def\ck{{\cal K}}
\def\cl{{\cal L}}
\def\cm{{\cal M}}
\def\cn{{\cal N}}
\def\co{{\cal O}}
\def\cp{{\cal P}}
\def\cq{{\cal Q}}
\def\car{{\cal R}}
\def\cs{{\cal S}}
\def\ct{{\cal T}}
\def\cu{{\cal U}}
\def\cv{{\cal V}}
\def\cw{{\cal W}}
\def\cx{{\cal X}}
\def\cy{{\cal Y}}
\def\cz{{\cal Z}}


\def\Sc#1{{\hbox{\sc #1}}}      
\def\Sf#1{{\hbox{\sf #1}}}      



\def\slpa{\slash{\pa}}                            
\def\slin{\SLLash{\in}}                                   
\def\bo{{\raise-.3ex\hbox{\large$\Box$}}}               
\def\cbo{\Sc [}                                         
\def\pa{\partial}                                       
\def\de{\nabla}                                         
\def\dell{\bigtriangledown}                             
\def\su{\sum}                                           
\def\pr{\prod}                                          
\def\iff{\leftrightarrow}                               
\def\conj{{\hbox{\large *}}}                            
\def\ltap{\raisebox{-.4ex}{\rlap{$\sim$}} \raisebox{.4ex}{$<$}}   
\def\gtap{\raisebox{-.4ex}{\rlap{$\sim$}} \raisebox{.4ex}{$>$}}   
\def\TH{{\raise.2ex\hbox{$\displaystyle \bigodot$}\mskip-4.7mu \llap H \;}}
\def\face{{\raise.2ex\hbox{$\displaystyle \bigodot$}\mskip-2.2mu \llap {$\ddot
        \smile$}}}                                      
\def\dg{\sp\dagger}                                     
\def\ddg{\sp\ddagger}                                   

\font\tenex=cmex10 scaled 1200


\def\sp#1{{}^{#1}}                              
\def\sb#1{{}_{#1}}                              
\def\oldsl#1{\rlap/#1}                          
\def\slash#1{\rlap{\hbox{$\mskip 1 mu /$}}#1}      
\def\Slash#1{\rlap{\hbox{$\mskip 3 mu /$}}#1}      
\def\SLash#1{\rlap{\hbox{$\mskip 4.5 mu /$}}#1}    
\def\SLLash#1{\rlap{\hbox{$\mskip 6 mu /$}}#1}      
\def\PMMM#1{\rlap{\hbox{$\mskip 2 mu | $}}#1}   %
\def\PMM#1{\rlap{\hbox{$\mskip 4 mu ~ \mid $}}#1}       %
\def\Tilde#1{\widetilde{#1}}                    
\def\Hat#1{\widehat{#1}}                        
\def\Bar#1{\overline{#1}}                       
\def\sbar#1{\stackrel{*}{\Bar{#1}}}             
\def\bra#1{\left\langle #1\right|}              
\def\ket#1{\left| #1\right\rangle}              
\def\VEV#1{\left\langle #1\right\rangle}        
\def\abs#1{\left| #1\right|}                    
\def\leftrightarrowfill{$\mathsurround=0pt \mathord\leftarrow \mkern-6mu
        \cleaders\hbox{$\mkern-2mu \mathord- \mkern-2mu$}\hfill
        \mkern-6mu \mathord\rightarrow$}
\def\dvec#1{\vbox{\ialign{##\crcr
        \leftrightarrowfill\crcr\noalign{\kern-1pt\nointerlineskip}
        $\hfil\displaystyle{#1}\hfil$\crcr}}}           
\def\dt#1{{\buildrel {\hbox{\LARGE .}} \over {#1}}}     
\def\dtt#1{{\buildrel \bullet \over {#1}}}              
\def\der#1{{\pa \over \pa {#1}}}                
\def\fder#1{{\d \over \d {#1}}}                 


\def\frac#1#2{{\textstyle{#1\over\vphantom2\smash{\raise.20ex
        \hbox{$\scriptstyle{#2}$}}}}}                   
\def\half{\frac12}                                        
\def\sfrac#1#2{{\vphantom1\smash{\lower.5ex\hbox{\small$#1$}}\over
        \vphantom1\smash{\raise.4ex\hbox{\small$#2$}}}} 
\def\bfrac#1#2{{\vphantom1\smash{\lower.5ex\hbox{$#1$}}\over
        \vphantom1\smash{\raise.3ex\hbox{$#2$}}}}       
\def\afrac#1#2{{\vphantom1\smash{\lower.5ex\hbox{$#1$}}\over#2}}    
\def\partder#1#2{{\partial #1\over\partial #2}}   
\def\parvar#1#2{{\d #1\over \d #2}}               
\def\secder#1#2#3{{\partial^2 #1\over\partial #2 \partial #3}}  
\def\on#1#2{\mathop{\null#2}\limits^{#1}}               
\def\bvec#1{\on\leftarrow{#1}}                  
\def\oover#1{\on\circ{#1}}                              

\def\[{\lfloor{\hskip 0.35pt}\!\!\!\lceil}
\def\]{\rfloor{\hskip 0.35pt}\!\!\!\rceil}
\def\Lag{{\cal L}}
\def\du#1#2{_{#1}{}^{#2}}
\def\ud#1#2{^{#1}{}_{#2}}
\def\dud#1#2#3{_{#1}{}^{#2}{}_{#3}}
\def\udu#1#2#3{^{#1}{}_{#2}{}^{#3}}
\def\calD{{\cal D}}
\def\calM{{\cal M}}

\def\szet{{${\scriptstyle \b}$}}
\def\ulA{{\un A}}
\def\ulM{{\underline M}}
\def\cdm{{\Sc D}_{--}}
\def\cdp{{\Sc D}_{++}}
\def\vTheta{\check\Theta}
\def\fracm#1#2{\hbox{\large{${\frac{{#1}}{{#2}}}$}}}
\def\ha{{\fracmm12}}
\def\tr{{\rm tr}}
\def\Tr{{\rm Tr}}
\def\itrema{$\ddot{\scriptstyle 1}$}
\def\ula{{\underline a}} \def\ulb{{\underline b}} \def\ulc{{\underline c}}
\def\uld{{\underline d}} \def\ule{{\underline e}} \def\ulf{{\underline f}}
\def\ulg{{\underline g}}
\def\items#1{\\ \item{[#1]}}
\def\ul{\underline}
\def\un{\underline}
\def\fracmm#1#2{{{#1}\over{#2}}}
\def\footnotew#1{\footnote{\hsize=6.5in {#1}}}
\def\low#1{{\raise -3pt\hbox{${\hskip 0.75pt}\!_{#1}$}}}

\def\Dot#1{\buildrel{_{_{\hskip 0.01in}\bullet}}\over{#1}}
\def\dt#1{\Dot{#1}}

\def\DDot#1{\buildrel{_{_{\hskip 0.01in}\bullet\bullet}}\over{#1}}
\def\ddt#1{\DDot{#1}}

\def\DDDot#1{\buildrel{_{_{\hskip 0.01in}\bullet\bullet\bullet}}\over{#1}}
\def\dddt#1{\DDDot{#1}}

\def\DDDDot#1{\buildrel{_{_{\hskip 
0.01in}\bullet\bullet\bullet\bullet}}\over{#1}}
\def\ddddt#1{\DDDDot{#1}}

\def\Tilde#1{{\widetilde{#1}}\hskip 0.015in}
\def\Hat#1{\widehat{#1}}


\newskip\humongous \humongous=0pt plus 1000pt minus 1000pt
\def\caja{\mathsurround=0pt}
\def\eqalign#1{\,\vcenter{\openup2\jot \caja
        \ialign{\strut \hfil$\displaystyle{##}$&$
        \displaystyle{{}##}$\hfil\crcr#1\crcr}}\,}
\newif\ifdtup
\def\panorama{\global\dtuptrue \openup2\jot \caja
        \everycr{\noalign{\ifdtup \global\dtupfalse
        \vskip-\lineskiplimit \vskip\normallineskiplimit
        \else \penalty\interdisplaylinepenalty \fi}}}
\def\li#1{\panorama \tabskip=\humongous                         
        \halign to\displaywidth{\hfil$\displaystyle{##}$
        \tabskip=0pt&$\displaystyle{{}##}$\hfil
        \tabskip=\humongous&\llap{$##$}\tabskip=0pt
        \crcr#1\crcr}}
\def\eqalignnotwo#1{\panorama \tabskip=\humongous
        \halign to\displaywidth{\hfil$\displaystyle{##}$
        \tabskip=0pt&$\displaystyle{{}##}$
        \tabskip=0pt&$\displaystyle{{}##}$\hfil
        \tabskip=\humongous&\llap{$##$}\tabskip=0pt
        \crcr#1\crcr}}


\def\eV{\,{\rm eV}}
\def\keV{\,{\rm keV}}
\def\MeV{\,{\rm MeV}}
\def\GeV{\,{\rm GeV}}
\def\TeV{\,{\rm TeV}}
\def\sv{\left<\sigma v\right>}
\def\({\left(}
\def\){\right)}
\def\cm{{\,\rm cm}}
\def\K{{\,\rm K}}
\def\kpc{{\,\rm kpc}}
\def\beq{\begin{equation}}
\def\eeq{\end{equation}}
\def\bea{\begin{eqnarray}}
\def\eea{\end{eqnarray}}


\newcommand{\be}{\begin{equation}}
\newcommand{\ee}{\end{equation}}
\newcommand{\nbe}{\begin{equation*}}
\newcommand{\nee}{\end{equation*}}

\newcommand{\fr}{\frac}
\newcommand{\lb}{\label}

\thispagestyle{empty}

{\hbox to\hsize{
\vbox{\noindent September 2024 \hfill IPMU24-0033} }}

\noindent  

\noindent
\vskip2.0cm
\begin{center}

{\large\bf Starobinsky inflation beyond the leading order}

\vglue.4in

Shunsuke Toyama~${}^{a,\&}$  and Sergei V. Ketov~${}^{a,b,c,\#,}$\footnote{The corresponding author} 
\vglue.3in

${}^a$~Department of Physics, Tokyo Metropolitan University,\\
1-1 Minami-ohsawa, Hachioji-shi, Tokyo 192-0397, Japan \\
${}^b$~Research School of High-Energy Physics, Tomsk Polytechnic University, \\
Tomsk 634028, Russian Federation\\
${}^c$~Kavli Institute for the Physics and Mathematics of the Universe (WPI),
\\The University of Tokyo Institutes for Advanced Study,  Kashiwa 277-8583, Japan\\

\vglue.2in

${}^{\&}$~toyama-shunsuke@ed.tmu.ac.jp, ${}^{\#}$~ketov@tmu.ac.jp
\end{center}

\vglue.4in

\begin{center}
{\Large\bf Abstract}  
\end{center}
\vglue.1in

\noindent  The Starobinsky model of cosmological inflation in four spacetime dimensions is reviewed with the emphasis on impact of quantum gravity corrections. As a specific example of the quantum corrections, the Grisaru-Zanon quartic curvature terms in the gravitational effective action of closed superstrings are chosen. Those quartic curvature terms are compared to the Bel-Robinson
tensor squared in a flat Friedman universe, and the upper bound on the effective string coupling constant is found by demanding unitarity
(causality)  and the absence of ghosts. It is found that the quantum corrections to the observables (tilts) of the cosmic microwave background radiation in the Starobinsky inflation may be of the same order of magnitude as the next-to-next-to-next classical contributions in the Starobinsky model with respect to the inverse powers of the e-folding number at the horizon crossing.

\newpage

\section{Introduction}

Ultra-violet (UV) completion of phenomenologically viable field-theoretical models of cosmological inflation is important because inflation is sensitive to high energy physics in the early Universe and, hence, quantum corrections to classical description of inflation may be important. The existence of the UV completion to a particular inflation model allows one to treat it as the effective field theory (EFT) originating from quantum gravity. The UV completion is even more relevant to large-field inflation models, while the Starobinsky inflation model \cite{Starobinsky:1980te} is one of them. Unfortunately, little is known about quantum gravity,
so UV-completion is often discussed in the framework of Swampland conjectures \cite{Palti:2019pca}, see also Refs.~\cite{Brinkmann:2023eph,Lust:2023zql,Ketov:2024klm} for the applications of the Swampland conjectures to the Starobinsky inflation.

String theory is a good candidate for the theory of quantum gravity, so it is natural to seek an UV completion of the Starobinsky model in string theory, see e.g., Ref.~\cite{Blumenhagen:2015qda} for the earlier attempts. This task turned out to be difficult because
the Starobinsky model is defined in four space-time dimensions, whereas string theory needs to be compactified from higher dimensions to four dimensions, while the gravitational low-energy EFT in string theory is subject to large ambiguities related to  field redefinitions of spacetime metric, see e.g., Ref.~\cite{Ketov:2000dy} for more. 

The Starobinsky model of inflation is based on the $R^2$ gravity and can be destabilized by the higher-order curvature
terms in the gravitational EFT if those terms have large coefficients. Inflation provides the mechanism for generation of cosmological perturbations, while the Starobinsky model is in excellent agreement with current measurements of the cosmic microwave background (CMB) radiation, so the higher order curvature terms must be subleading during inflation. Nevertheless, it makes sense to investigate robustness of the Starobsinky inflation against specific quantum gravity corrections derived from superstring theory. In the case of closed (type II) superstrings, the leading $(\alpha')^3$-correction beyond the Einstein-Hilbert term is given by the terms quartic in the space-time curvature, which were first derived by Grisaru and Zanon in 1985 from the vanishing four-loop renormalization group beta-function of the supersymmetric non-linear sigma model in two dimensions, describing propagation of a test superstring in the gravitational background \cite{Grisaru:1986vi}. The same quartic curvature terms arise from M-theory \cite{Green:1997as,Blumenhagen:2024ydy} after dimensional reduction down to four dimensions. The dependence of the gravitational EFT in string theory upon the Ricci scalar curvature (and the Ricci tensor also) is known to be ambiguous, see Ref.~\cite{Ketov:2000dy} for a review, because  the perturbative string theory is consistently defined only on Ricci-flat backgrounds.

In this paper, we combined the Starobinsky model with the Grisaru-Zanon (GZ) quantum gravity (superstring) correction, called the Starobinsky-Grisaru-Zanon (SGZ) gravity, that is the generalization of  the Einstein-Grisaru-Zanon gravity introduced in 
Ref.~\cite{CamposDelgado:2024jst}. It allowed us not only derive the restrictions on the effective superstring coupling constant in front
of the quartic curvature terms but also compare the contributions of those quantum corrections to the subleading terms (beyond the
leading order with respect to the e-folding number) in the inflationary (CMB) observables such as cosmological tilts and their running, as well as derive the leading quantum gravity (superstring) corrections to the Starobinsky solution.

The paper is organized as follows. In Sec.~2 the Starobinsky model of inflation is reviewed both in the original (Jordan) frame and in
the Einstein (quintessence) frame. We do not follow historical developments but introduce the Starobinsky model from the modern perspective. In Sec.~3 we define the SGZ gravity in the perturbative setup with respect to the GZ term, and compare it to the similar but different Starobinsky-Bel-Robinson (SBR) gravity \cite{Ketov:2022lhx,CamposDelgado:2022sgc,Ketov:2022zhp} also having the quartic curvature terms in its action. In Sec.~4 we derive the leading quantum corrections to the inflationary dynamics due to the GZ term in the Jordan frame, in the first order with respect to its effective (string) coupling constant. In Sec.~5 we find the upper limits on the effective string coupling constant by demanding the absence of ghosts, unitarity and causality. The leading quantum corrections to the CMB observables are derived and compared against the classical subleading contributions in Sec.~6. We conclude in Sec.~7.

\section{Review of Starobinsky inflation}

The Starobinsky model of inflation is the generally covariant and nonperturbative extension of the Einstein-Hilbert (EH) gravity theory by the term quadratic in the Ricci scalar curvature $R$. All the curvature-dependent terms beyond the EH one are irrelevant in the Solar system, while they may also be negligible during reheating after inflation in the weak-gravity regime. However, it is not the case during inflation in the high curvature regime where the $R^2$ term is the leading contribution (see below).

The Starobinsky model is the particular case of modified $F(R)$ gravity, and it is geometrical because only gravitational interactions are used. A modified gravity action has the higher derivatives and generically suffers from Ostrogradsky instabilities and ghosts. However, in the most general modified gravity action, whose Lagrangian is quadratic in the spacetime curvature, the  only ghost-free term is just given by $R^2$ with a positive coefficient, which leads  to the Starobinsky model with the action 
\be  S_{\rm Star.} = \alpha \int \mathrm{d}^4x\sqrt{-g} R^2 + \fracmm{M^2_{\rm Pl}}{2}\int \mathrm{d}^4x\sqrt{-g} R~,\quad \alpha \equiv \fracmm{M^2_{\rm Pl}}{12M^2} ~, \lb{star}
\ee
having the only parameter $\a$ or $M$, where $M_{\rm Pl}=1/\sqrt{8\p G_{\rm N}}\approx 2.4\times 10^{18}$ GeV, the spacetime signature is  $(-,+,+,+,)$ and the natural units are used, $\hbar=c=1$. The first term in this action is scale invariant with the dimensionless parameter
$\a$.

The origin of the $R^2$ term was originally proposed due to contributions of quantized matter fields in the EH gravity 
\cite{Starobinsky:1980te}. However, because the EH term is subleading during inflation, we adopt the
opposite interpretation, namely, with the EH term being originated from the scale-invariant gravity. For instance, when starting from 
the scale-invariant action for gravity and a scalar field $\phi$ as
\cite{Cooper:1981byv,Buchbinder:1986wk,Einhorn:2015lzy}
\be S[g_{\m\n},\phi] =  \int \mathrm{d}^4x\sqrt{-g} \left[ \alpha R^2 + \x \phi^2 R -\ha (\pa\phi)^2 -\l \phi^4\right]~,
\ee
one finds that it can undergo a phase transition (called dimensional transmutation) due to quantum corrections, known as the 
 Coleman-Weinberg mechanism of spontaneous symmetry breaking \cite{Coleman:1973jx}. It leads to the massive scalar field $\phi$ that may  be identified with dilaton or Higgs field having  a non-vanishing vacuum expectation value (VEV)  in the effective action, as can be demonstrated in the one-loop perturbation theory \cite{Buchbinder:1986wk,Einhorn:2015lzy}. As a result, both the Planck mass and the EH term are generated with 
\be \ha M^2_{\rm Pl} = \x \VEV{\phi}^2~,
\ee
though this cannot be considered as the UV-completion of the Starobinsky gravity.

The metric of a flat Friedman universe is given by 
\be ds^2=-dt^2+a^2\left(dx_1^2+dx_2^2+dx_3^2\right)~. \lb{flatF}
\ee
Then the action (2) leads to equations of motion in the form
\be   \lb{stareom}
2H\ddot{H} - \left(\dot{H}\right)^2 + H^2\left(6\dot{H} + M^2\right)=0~,\quad H=\dot{a}/a~,
\ee
known as the Starobinsky equation in the literature, where the dots stand for the time derivatives and $H(t)$ is Hubble function.

When searching for a solution to the Starobinsky equation in the form of left Painlev\'e series, 
$H(t)=\sum^{k=p}_{k=-\infty}c_k(t_0-t)^k$, one finds the Hubble function (see e.g., Ref.~\cite{Ketov:2022zhp})
\be \lb{stars}
\begin{split}
\fracmm{H(t)}{M}  & =  \fracmm{M}{6}(t_0-t)+\fracmm{1}{6M(t_0-t)} - \fracmm{4}{9M^3(t_0-t)^3}+
\fracmm{146}{45M^5(t_0-t)^5}  \\
& {} -\fracmm{11752}{315 M^7 (t_0-t)^7} + {\cal O} \left(M^{-9}(t_0-t)^{-9}\right)
\end{split}
\ee
valid for $M(t_0-t)>1$. This special solution is an attractor, while $R=12H^2+6\dot{H}$.

In the slow-roll (SR) approximation defined by $\abs{\ddot{H}} \ll \abs{H\dot{H}}$ and  $\abs{\dot{H}} \ll H^2$, one gets the leading
term in the Starobinsky solution as 
\be 
 H(t) \approx \left (\fracmm{M^2}{6}\right) (t_0-t)
\ee
that is entirely due to the $R^2$-term in the action. The attractor solution spontaneously breaks the scale invariance of the $R^2$-gravity and, therefore, implies the existence of the Nambu-Goldstone boson (called scalaron) that is the physical excitation of the higher-derivative gravity. It can be made manifest by rewriting the Starobinsky action into the quintessence form after the field redefinition (or Legendre-Weyl transform) \cite{Maeda:1988ab}
\begin{equation} 
  \varphi =  \sqrt{ \fracmm{3}{2}} M_{\rm Pl}\ln F'(\c)   \quad {\rm and}\quad g_{\m\n}\to \fracmm{2}{M^2_{\rm Pl}}F'(\chi) g_{\m\n}~,
  \quad \chi=R~.
  \ee
It yields
\be S[g_{\m\n},\varphi]  = \fracmm{M^2_{\rm Pl}}{2}\int \mathrm{d}^4x\sqrt{-g} R 
 - \int \mathrm{d}^4x \sqrt{-g} \left[ \frac{1}{2}g^{\m\n}\pa_{\m}\varphi\pa_{\n}\varphi
 + V(\varphi)\right]~,\lb{quint}
\ee
in terms of the canonical inflaton $\varphi$ with the scalar potential 
\be \lb{starp}
V(\varphi) = \fracmm{3}{4} M^2_{\rm Pl}M^2\left[ 1- \exp\left(-\sqrt{\frac{2}{3}}\varphi/M_{\rm Pl}\right)\right]^2~.
\ee
This potential has the infinite plateau (for the large $\phi$-field values of the order $M_{\rm Pl}$ and beyond)  that implies the approximate shift symmetry of the inflaton field as the consequence of the scale invariance of the $R^2$ gravity or due to the approximate scale invariance of the action (2) in the large-curvature regime. The potential (\ref{starp}) also has the positive "cosmological constant" given by the first term in the square brackets, induced by the $R^2$ term in the action (\ref{star}), which can be physically interpreted as the energy driving inflation. The scale of inflation is determined by the parameter $M$ that is identified with the inflaton mass. The universality class of inflationary models is determined by the critical parameter $\sqrt{2/3}$ in the exponential term \cite{Ketov:2021fww}.

The equivalent actions (\ref{star}) and (\ref{quint}) are usually referred to the Jordan frame and the Einstein frame, respectively. The approximate shift symmetry of the potential (\ref{starp}) is the consequence of the approximate scale invariance of the $R^2$ gravity, which requires the presence of the $R^2$ term in any viable model of inflation based on modified $F(R)$-gravity. It becomes even more transparent by using the inverse transformation from the Einstein frame to the Jordan frame, having the parametric form \cite{Ketov:2014jta}
\be
R = \left(  \fracmm{\sqrt{6}}{M_{\rm Pl}}
    \fracmm{d V}{d \varphi} + \fracmm{4V}{M^2_{\rm Pl}} \right) e^{ \sqrt{\frac{2}{3}} 
  \varphi/M_{\rm Pl}}~,    \quad F= \left(  \fracmm{\sqrt{6}}{M_{\rm Pl}}
 \fracmm{d V}{d \varphi} + \fracmm{2V}{M^2_{\rm Pl}} \right) e^{ 2 \sqrt{\frac{2}{3}} \varphi/M_{\rm Pl}}~.
\ee
As is clear from these equations, in the SR approximation (chaotic inflation) the first term in the brackets is much less than the second term, which immediately implies $F(R)\sim R^2$. 

The gravitational EFT during inflation does not have to be limited to the terms given in Eq.~(\ref{star})  but should also include 
the higher-order curvature terms. Those terms eliminate the infinite plateau in the inflaton potential (\ref{starp}).  The fact that  the Starobinsky model of inflation is in excellent agreement with the  current CMB measurements (see below) implies that
those terms do not destabilize the Starobinsky inflation, which put restrictions on their contributions.

It is convenient to use the e-foldings number $N$ instead of time $t$, which are related by
\be 
N=\int^{t_0}_{t} H(\tilde{t}) d\tilde{t}~,
\ee
and the co-moving wavenumber $k=2\pi/\lambda$ related to $N(t)$ by the equation $d\ln k =-dN$~.

The SR (running) parameters in the Einstein frame are defined by
\be
\ve_{\rm sr}(\varphi) = \fracmm{M^2_{\rm Pl}}{2}\left( \fracmm{V'}{V}\right)^2 \quad {\rm and} \quad 
\eta_{\rm sr}(\varphi) = M^2_{\rm Pl} \left( \fracmm{V''}{V}\right)
\ee
in terms of the quintessence scalar potential $V$, where the primes denote the derivatives with respect to $\varphi$. In  the Jordan frame,
one uses the Hubble flow functions,
\be \lb{Hflow}
\epsilon_{H} = -\fracmm{\dot{H}}{H^{2}}~,\quad 
	\eta_{H} = \epsilon_{H} - \fracmm{\dot{\epsilon}_{H}}{2\epsilon_{H} H}~~.
\ee

The amplitude of scalar perturbations at the horizon crossing with the pivot scale $k_*=0.05~{\rm Mpc}^{-1}$  is known from CMB
measurements (called WMAP normalization) as
\be 
A_s= \fracmm{V_*^3}{12\p^2 M^6_{\rm Pl}({V_*}')^2}=\fracmm{3M^2}{8\p^2M^2_{\rm Pl}}\sinh^4\left(
\fracmm{\varphi_*}{\sqrt{6}M_{\rm Pl}}\right)\approx 2 \cdot 10^{-9}~,
\ee
where subscript (*) refers to the CMB pivot scale, in the case of Starobinsky inflation. This allows us to fix the only parameter $M$ (or $\a$) and  the scale of inflation, $H_{\rm inf.}$, in the Starobinsky model as 
\be
\fracmm{M}{M_{\rm Pl}}\approx   {\cal O}(10^{-5}), \quad \a\approx {\cal O}(10^9), \quad H\approx {\cal O}(10^{14})~{\rm GeV},\quad 
\fracmm{R}{M^2_{\rm Pl}} \approx \fracmm{12H^2}{M^2_{\rm Pl}}\approx 10^{-7}.
\ee
It is worth mentioning here that the higher-order curvature terms in the gravitational EFT beyond the Starobinsky model are given by power series with respect to  $H^2/M^2_{\rm Pl}\sim 10^{-8}$, so they must be sub-leading during inflation unless they have very large coefficients. It is also worth noticing that the large value of $\alpha$ required by CMB does not speak in favor of generating 
the $R^2$-term by quantum matter contributions because a single quantized matter field contributes in the 1-loop approximation about
$10^{-3}$ to the $\alpha$-parameter, so one needs about $10^{12}$ quantized matter fields in order to achieve the desired result.

The primordial spectrum $P_{\z}(k)$ of 3-dimensional scalar (density) perturbations $\z(x)$ in a flat Friedman universe is defined by the 2-point correlation function 
\be \lb{defpsp}
\VEV{ \fracmm{\d\z(x)}{\z} \fracmm{\d\z(y)}{\z}} =\int\fracmm{d^3k}{k^3} e^{ik\cdot (x-y)}\fracmm{P_{\z}(k)}{P_0}~~,
\ee
where $k=2\p/\l$ is the co-moving number. Similarly, one defines the primordial spectrum $P_t(k)$ of tensor perturbations, 
see e.g., Ref.~\cite{Ketov:2021fww} for more details The scale $k$ is simply related to the e-folds number $N$ via $N=-\int^k d\tilde{k}/\tilde{k}$.
The power spectra coincide with the corresponding amplitudes $A_s(k)$ and $A_t(k)$, respectively. 

Given the power spectra $P_{\z}(k)$ and $P_t(k)$,  one defines the scalar tilt $n_s(k)$, its running parameter $\a_s(k)$, the tensor tilt $n_t(k)$ and its running parameter $\a_t(k)$ (all dimensionless) as
\be \lb{tilts}
n_s = 1 + \fracmm{d\ln P_{\z}(k)}{d\ln k}~, \quad \a_s =  \fracmm{d^2\ln P_{\z}(k)}{(d\ln k)^2}~,\quad n_t = \fracmm{d\ln P_{t}(k)}{d\ln k}~, 
 \quad \a_t =  \fracmm{d^2\ln P_t(k)}{(d\ln k)^2}~~,
 \ee
 as well as the tensor-to-scalar ratio 
\be \lb{ts}   r(k) = \fracmm{P_t}{P_\z}=8\abs{n_t}~.
\ee

The Starobinsky model  gives simple predictions for the cosmological tilts of the scalar and tensor power spectra {\it in the leading order} with respect to the e-folds $N_*$ evaluated when perturbations left the horizon (at the horizon crossing)  as \cite{Mukhanov:1981xt,Mukhanov:1990me}
\be 
n_s\approx 1- \fracmm{2}{N_*}~~,\quad \a_s\approx -\fracmm{2}{N^2_*}~~,\quad \a_t\approx -\fracmm{3}{N^3_*}~,\quad 
r\approx \fracmm{12}{N^2_*}~~. \lb{strtilts}
\ee
Therefore, tensor perturbations are suppressed with respect to scalar perturbations by the extra factor of $N^{-1}_*$, whose value can be
estimated by comparing those predictions with CMB measurements \cite{BICEP:2021xfz,Tristram:2021tvh}~,
\be  \lb{cmb}
n_s\approx 0.9649\pm 0.0042~(68\% {\rm CL}) \quad {\rm and} \quad r< 0.032~(95\% {\rm CL})~,
\ee
that fit the Starobinsky model predictions for 
\be \lb{Nstar}
N_*=56 \pm 8~.
\ee
This prediction for the duration of inflation agrees with our calculations in the Jordan frame, based on the solution (\ref{stars}).
The corresponding times for the end and the beginning of inflation are $M(t_0-t_{\rm end})\approx 2.5$ and $M(t_0-t_{\rm start})\approx 27.7$, respectively. 

In particular, excluding $N_*$ from Eqs.~(\ref{strtilts}) yields the sharp prediction of the Starobinsky model  for the tensor-to-scalar ratio as
\be r\approx 3(1-n_s)^2~. \lb{starrel}
\ee

The Starobinsky inflation does not exclude the higher-order curvature terms in the action (\ref{star}), though it implies that those terms should be subleading during inflation, being suppressed by the powers of $H^2/M^2_{\rm Pl}\sim 10^{-8}$. The Starobinsky model is sensitive to quantum (UV) corrections because of its high inflation scale and the inflaton field values near the Planck scale during inflation. Hence, it is important to determine its UV-cutoff $\L_{\rm UV}$ of the Starobinsky model by studying scaling of scattering amplitudes with respect to energy, $E/\L_{\rm UV}$.
A careful calculation yields \cite{Hertzberg:2010dc}
\be 
\Lambda_{\rm UV}=M_{\rm Pl}~.
\ee
Therefore, the predictions of the Starobinsky model for inflation and CMB make sense and the model itself can be considered as a trustable effective field theory after decoupling of heavy modes expected at the Planck scale \cite{Ketov:2024klm}.

\section{Starobinsky-Grisaru-Zanon (SGZ) gravity}

The SGZ action is defined by
\begin{equation} \lb{sgz}
    S_{\rm SGZ}[g]=\fracmm{M_{\text{Pl}}^2}{2}\int d^4x \sqrt{-g}\,\left(R+ \fracmm{1}{6M^2}R^2 -\fracmm{72\g}{M^6}Z\right), 
\end{equation}
where we have added the GZ (quantum) superstring correction \cite{Grisaru:1986vi} 
\begin{equation}
    72Z=\left(R^{\mu\rho\sigma\nu}R_{\lambda\rho\sigma\tau}+\frac{1}{2}R^{\mu\nu\rho\sigma}R_{\lambda\tau\rho\sigma}\right)R_{\mu}^{\,\,\,\alpha\beta\lambda}R^{\tau}_{\,\,\,\alpha\beta\nu}
\end{equation}
to the Starobinsky action (\ref{star}) with the new dimensionless coupling constant $\g>0$~\footnote{The parameter $\g$ introduced in Ref.~\cite{CamposDelgado:2024jst} was rescaled here by the factor $(M_{\rm Pl}/M)^6$.}.

The value of $\gamma$ cannot be calculated from string theory because the action (\ref{sgz}) is in four space-time dimensions, so $\gamma$ depends upon compactification from ten to four dimensions and the unknown vacuum expectation value of the string dilaton.

The SGZ gravity (\ref{sgz}) is different from the SBR gravity defined by the action \cite{Ketov:2022lhx,Ketov:2022zhp} 
\be
S_{\rm SBR}  =  \fracmm{M_{\rm Pl}^2}{2}\int d^{4}x\,\sqrt{-g}\left( R  +\fracmm{1}{6M^2}R^2 
-\fracmm{\b}{8M^6}  T^2\right)~,
\label{ebr}
\ee
where $\b$  is another dimensionless coupling constant, the $T^2$ stands for the Bel-Robinson (BR) tensor squared,
$T^2\equiv T_{\r\n\l\m}T ^{\r\n\l\m}$, with 
\be
T^{\r \n\l\m} \equiv R^{\r \s \h\l}R\udu{\n}{\s \h}{\m} +{}^*R^{\r \s \h\l}{}^*R\udu{\n}{\s \h}{\m} =R^{\r \s \h\l}R\udu{\n}{\s\h}{\m}
 +R^{\r \s \h\m}R\udu{\n}{\s \h}{\l}-\ha g^{\r\n}R^{\s \h \x\l} R\du{\s \h \x}{\m}~,
\label{belr}
\ee
and the star denoting the Hodge dual tensor in four dimensions. 

The BR tensor was introduced by Bel and Robinson \cite{Bel:1959uwe,Robinson:1997} by analogy with the energy-momentum tensor of Maxwell theory of electromagnetism,
\be T^{\rm Maxwell}_{\m\n}=F_{\m\r}F\du{\n}{\r}+{}^*F_{\m\r}{}^*F\du{\n}{\r}~~,\quad 
F_{\m\n}=\pa_{\m}A_{\n}-\pa_{\n}A_{\m}\lb{max}~~.
\ee
There is an  identity \cite{Deser:1999jw,Iihoshi:2007vv}
\be
T^2=  -\frac{1}{4}({}^*\!R_{\m\n\l\r}{}^*\!R^{\m\n\l\r})^2 
+\frac{1}{4}({}^*R_{\m\n\l\r}R^{\m\n\l\r})^2 
 = \frac{1}{4}(P_4^2-E^2_4) =\frac{1}{4}(P_4+E_4)(P_4-E_4) \label{id2}
\ee
relating the BR tensor squared to the Euler and Pontryagin topological densities in four dimensions, $E_4$ and $P_4$, respectively. 
There is another well-known identity
\be \lb{gb}
E_4 = {\cal G}_{\rm GB} \equiv R^{\m\n\l\r}R_{\m\n\l\r} - 4R^{\m\n}R_{\m\n} + R^2
\ee
that relates the Euler density to the Gauss-Bonnet invariant ${\cal G}_{\rm GB}$  in four dimensions.

We consider the first two terms in the SGZ and SBR actions nonperturbatively, but the last (quartic) curvature (BR or GZ) terms 
only perturbatively, in the first order with respect to the coupling constants. Therefore, no Ostrogradski ghosts arise. The difference between the GZ and BR terms in four dimensions, in the context  of superstrings/M-theory, was first noticed in Ref.~\cite{Moura:2007ks}. 

In a flat Friedman universe (\ref{flatF}) we find 
\be Z= H^8  + 2H^6\dot{H} +\fracm{11}{6} H^4\dot{H}^2 +\fracm{2}{3}H^2\dot{H}^3 + \fracm{1}{12} \dot{H}^4~,
\ee
whereas
\be 
\fracm{1}{144} T^2 = H^8 + 2 H^6\dot{H} + H^4 \dot{H}^2~,
\ee
which makes the difference manifest, though the first two leading terms (relevant to the SR approximation) are the same.

\section{GZ quantum corrections to Starobinsky inflation}

The SGZ gravity equation of motion in a flat Friedman universe is given by
\be \lb{eom}
\eqalign{
& m^{6} H^{2} + 6m^{4} H^{2} \dot{H} + 2m^{4} H \ddot{H} -m^{4} \dot{H}^{2} \cr
& - 12 \gamma H^{8} + 132 \gamma H^{6} \dot{H} + 44 \gamma H^{5} \ddot{H} + 138 \gamma H^{4} \dot{H}^{2} \cr
& + 48 \gamma H^{3} \dot{H} \ddot{H} + 28 \gamma H^{2} \dot{H}^{3} + 12 \gamma H \dot{H}^{2} \ddot{H} - 3 \gamma \dot{H}^{4}=0\cr}
\ee
that extends the Starobinsky equation (\ref{stareom}) by the $\gamma$-dependent terms. A solution to this equation in the first order
with respect to the (small) $\gamma$-parameter, similarly to Eq.~(\ref{stars}) reads
\be H(t) = H_0(t) +\g H_1(t)~,
\ee
where
\be \eqalign{
\fracmm{H_1(t)}{M}= & -\fracmm{1}{163296}M^7(t_0-t)^7-\fracmm{2}{2835}M^5(t_0-t)^5-\fracmm{391}{90720}M^3(t_0-t)^3-\fracmm{9061}{306180}M(t_0-t) \cr
& -\fracmm{127}{7776M(t_0-t)}-\fracmm{1931203}{5358150M^3(t_0-t)^3}+{\cal O}(M^{-5}(t_0-t)^{-5})~, \cr} \lb{gcor}
\ee
and the leading contribution during inflation comes from the 2nd term above.

Accordingly, the scalar curvature is given by
\be  \eqalign{
\fracmm{R}{M^2}= & ~~\fracmm{M^2 (t_0-t)^2}{3}-\fracmm{1}{3}-\fracmm{4}{9 M^2(t_0-t)^2}+\fracmm{16}{5 M^4 (t_0-t)^4}-\fracmm{6908}{189 M^6 (t_0-t)^6} \cr
&  +\gamma \left[-\fracmm{M^8 (t_0-t)^8}{40824}-\fracmm{151 M^6 (t_0-t)^6}{58320}+\fracmm{143 M^4 (t_0-t)^4}{122472}-\fracmm{4163 M^2 (t_0-t)^2}{81648} \right. \cr
&  -\fracmm{68713}{7144200} -\fracmm{5109281}{5143824 M^2(t_0-t)^2}+\fracmm{17584432631}{964467000 M^4 (t_0-t)^4} \cr
& \left.  -\fracmm{75802186291}{300056400 M^6 (t_0-t)^6}+{\cal O}(M^{-8}(t_0-t)^{-8})\right]~,\cr}
\ee
where the leading $\g$-dependent contribution during inflation is due to the 2nd term in the square brackets also.

The Hubble flow functions (\ref{Hflow}) are given by 
\be \eqalign{
\epsilon_H= & ~~\fracmm{6}{M^{2} (t_0-t)^{2}} - \fracmm{6}{M^{4} (t_0-t)^{4}} + \fracmm{48}{M^{6} (t_0-t)^{6}}  \cr
&  +\gamma \left[- \fracmm{5 M^{4} (t_0-t)^{4}}{4536} - \fracmm{869 M^{2} (t_0-t)^{2}}{11340} - \fracmm{4591}{22680}
+ \fracmm{38051}{34020 M^{2} (t_0-t)^{2}} \right. \cr
& \left.  - \fracmm{1385}{648 M^{4} (t_0-t)^{4}} + \fracmm{51649511}{396900 M^{6} (t_0-t)^{6}} \right]+
{\cal O}(M^{-8}(t_0-t)^{-8})  \cr} \lb{egam}
\ee
and
\be \eqalign{
\eta_H= & ~~\fracmm{6}{M^{4} (t_0-t)^{4}} - \fracmm{96}{M^{6} (t_0-t)^{6}} \cr 
& +\gamma \left[- \fracmm{5 M^{4} (t_0-t)^{4}}{1512} - \fracmm{869 M^{2} (t_0-t)^{2}}{5670} - \fracmm{4591}{22680} \right. \cr
& \left. + \fracmm{1385}{648 M^{4} (t_0-t)^{4}} - \fracmm{51649511}{198450 M^{6} (t_0-t)^{6}}\right] +{\cal O}(M^{-8}(t_0-t)^{-8})~,
\cr} \lb{etagam}
\ee
where the leading $\g$-dependent contributions during inflation are due to the 2nd terms in the square brackets too.
 
\section{The upper bounds on $\gamma$}

Some modified gravity models of inflation can be described by the effective function $F(H^2)$ entering equations of motion in a flat
Friedman universe~\cite{Cano:2020oaa}. Our equations of motion (\ref{eom}) are not of that type because they include the higher time derivatives of the Hubble function, but they fall into that type in the SR approximation, see Ref.~\cite{Ketov:2022zhp} for applications to the SBR gravity theory. In the case of SGZ gravity, we find
\be 
R\left(\fracmm{R}{12}-H^2\right)-H\dot{R}=3M^2\left(H^2-\fracmm{12\gamma H^8}{M^6}+\fracmm{22\gamma H^6}{M^4}\right)\equiv 3M^2F(H^2)~,\ee
so we have 
\be
F(H^2)=H^2-\fracmm{22\gamma}{M^4}\left(H^2\right){}^3-\fracmm{12\gamma}{m^6}\left(H^2\right){}^4~.
\ee
The derivatives of $F(H^2)$ with respect to $H^2$ are as follows:
\be \lb{Ffder}
F'(H^2) \equiv \fracmm{dF}{d(H^2)}=1-66\gamma\left(\fracmm{H}{M}\right)^4-48\gamma\left(\fracmm{H}{M}\right)^6
\ee
and
\be
F''(H^2)=-\fracmm{132\gamma}{M^4}H^2-\fracmm{144\gamma}{M^6}H^4~.
\ee

The effective Newton constant in the higher-derivative gravity theories studied in Ref.~\cite{Cano:2020oaa}
must obey the condition 
\begin{equation}
G_{\mathrm{eff.}}=\fracmm{1}{8\pi M_{\mathrm{Pl}}^2[F'(H^2)+4(H^2/M^2)]}>0~,
\ee
that implies
\be
F'(H^2)+4\fracmm{H^2}{M^2}>0 \quad {\rm or} \quad 
1-66\gamma h^4-48\gamma h^6+4 h^2>0~,
\ee
where $h=H/M$. The maximal value of $h$ in the Starobinsky inflation is $h_{\rm max.} \approx 4.6$, which implies
\be \lb{bound1}
\gamma<1.74\times 10^{-4}~.
\ee

Another condition proposed in Ref.~\cite{Cano:2020oaa} from demanding the absence of negative energy fluxes (or unitarity and causality
constraints) reads
\begin{equation}
-4\leq \fracmm{210H^2 F''(H^2)}{F'(H^2)+4(H^2/M^2)}\leq 4~,
\end{equation}
that in our case is given by
\begin{equation}
-1\leq \fracmm{210(-33\gamma h^4-38\gamma h^6)}{1-66\gamma h^4-48\gamma h^6+4 h^2}\leq 1~.
\end{equation}
It implies 
\be
\gamma\leq\fracmm{1+4h^2}{12h^4(634h^2+583)}~~,
\ee
and, therefore, the upper bound 
\be \lb{bound2}
\gamma\leq 1.12\times10^{-6}~.
\ee
Both bounds (\ref{bound1}) and (\ref{bound2}) are slightly stronger than those found in Ref.~\cite{Ketov:2022zhp} 

\section{Quantum corrections versus classical corrections to CMB observables beyond the leading order}

It is instructive to compare contributions of the GZ quantum correction to the CMB observables against the classical
contributions beyond the leading order given by Eq.~(\ref{strtilts}) within the possible range of $N_*$ in Eq.~(\ref{Nstar}).

The subleading terms for the predicted CMB observables in the Starobinsky inflation are given by \cite{Bianchi:2024qyp,Saburov:2022gbj}
\be \lb{subs}
\eqalign{
n_s = & ~~1- \fracmm{2}{N_*} + \fracmm{2.4}{N^2_*} - \fracmm{\ln(2N_*)}{6N^2_*} + {\cal O}\left( \fracmm{\ln 2N_*}{N^3_*}\right)~,\cr
\a_s = &-\fracmm{2}{N^2_*} +  {\cal O}\left( \fracmm{\ln 2N_*}{N^3_*}\right)~,\cr
r = & ~~\fracmm{12}{N^2_*} + \fracmm{2\ln (2N_*)}{N^3_*} - \fracmm{56.76}{N^3_*} + {\cal O}\left( N^{-4}_*\right)~,\cr
\a_t = & -\fracmm{3}{N^3_*}  +  {\cal O}\left( N^{-4}_*\right)~,\cr}
\ee
where $\ln 2N_*/N^3_* <  4.1\cdot 10^{-5}$ and $N^{-4}_* < 2\cdot 10^{-7}$. The subleading contribution to the scalar tilt $n_s$ in the 3rd term above can increase the $n_s$-value by $1.0\cdot 10^{-3}$, while the subleading contribution to the tensor-to-scalar-ratio $r$, given by
the 3rd term above, can decrease the $r$-value by $5\cdot 10^{-4}$. All the subleading contributions are within the observational errors  given in Eq.~(\ref{cmb}).

On the other hand, when using Eqs.~(\ref{egam}) and (\ref{etagam}) in the first order with respect to the string parameter $\g$ and taking their first-order contributions to the scalar tilt $n_s$ and the tensor-to-scalar-ratio $r$ at the horizon crossing,
\be 
n_s\approx 1 -4 \e_H +2\eta_H~, \quad r=16\e_H~,
\ee
with the maximal value of $\g$ from Eq.~(\ref{bound2}), we get the quantum contributions to the $n_s$ and $r$ up to $+2.5\cdot 10^{-4}$
and $-4.9\cdot 10^{-5}$, respectively. 

Therefore, the quantum contributions to the CMB tilts are smaller than the subleading terms proportional to $N^{-2}_*$ by one order
of magnitude but may be of the same order of magnitude as the classical $N^{-3}_*$ contributions. The same conclusion also applies to
the running parameters $\a_s$ and $\a_t$.

\section{Conclusion}

The upcoming CMB measurements by the CORE Collaboration \cite{CORE:2016ymi}, S4 Collaboration \cite{CMB-S4:2020lpa}, 
 LiteBIRD Collaboration \cite{Paoletti:2022kij}, NASA PICO Collaboration \cite{NASAPICO:2019thw}, the Simons Observatory survey \cite{SimonsObservatory:2018koc} and EUCLID Collaboration  \cite{Euclid:2021qvm} are expected to probe the tensor-to-scalar ratio $r$ in the range of $10^{-3}$ and improve the precision value of $n_s$.
 
 Should those measurements confirm Eq.~(\ref{starrel}), it would be a triumph of the Starobinsky model of inflation. Should the  predicted  relation (\ref{starrel}) be ruled out, the question would arise about the origin of disagreement. If the disagreement will be significant, it would
 rule out the Starobinsky inflation. If, however, the disagreement will be small (say, within one order of magnitude), one may expect that due to the sub-leading corrections to the leading order predictions in Eq.~(\ref{strtilts}). Then another question about the origin of those small corrections would arise. The latter may be due to the sub-leading terms in the classical Starobinsky model or due to quantum gravity corrections to the gravitational EFT, or  they have a very different origin, say, due to reheating or new physics, see e.g.,
 Ref.~\cite{Aldabergenov:2018qhs}.
 
 The main new result of this paper is about possible superstring (as quantum gravity) corrections that may be of the same size as the next-to-next-to-next classical corrections (in the $N^{-3}_*$-terms). Of course, the value of the effective string coupling $\g$ may be much lower than the estimate found in Eq.~(\ref{bound2}), which may reduce the size of quantum corrections even further.

\section*{Acknowledgements}

One of the authors (SVK) is grateful to Ignatios Antoniadis, Eugenio Bianchi, Norma Borstnik, Gia Dvali, Maxim Khlopov, Elias Kiritsis and Holger Nielsen for discussions.  SVK was supported by Tokyo Metropolitan University, the Japanese Society for Promotion of Science under the grant No.~22K03624, the World Premier International Research Center Initiative (MEXT, Japan), and the Tomsk Polytechnic University development program 
Priority-2030-NIP/EB-004-375-2024.

\bibliographystyle{utphys}
\bibliography{references}

\end{document}